\documentclass[pra,aps]{revtex4}

\input psfig.sty

\newcommand{\ra}{\rangle} 
\newcommand{\ba}{\begin{eqnarray}} 
\newcommand{\ea}{\end{eqnarray}} 
\newcommand{\be}{\begin{equation}} 
\newcommand{\ee}{\end{equation}} 
\newcommand{\bea}{\begin{eqnarray}} 
\newcommand{\eea}{\end{eqnarray}}

\begin{document} 
\def\CC{{\rm\kern.24em \vrule width.04em height1.46ex depth-.07ex \kern-.30em 
C}}  

\title{Encoded Universality for Generalized Anisotropic Exchange Hamiltonians} 
 
\author{Jiri Vala and K. Birgitta Whaley} 
 
\affiliation{Department of Chemistry, University of California, Berkeley, CA 
94720} 
 
\begin{abstract}
We derive an encoded universality representation for a generalized anisotropic
exchange Hamiltonian that contains cross-product terms in addition to the 
usual two-particle exchange terms.  The recently developed algebraic approach is used
to show that the minimal universality-generating encodings of one logical qubit are
based on three physical qubits.  We show how to generate both single- and two-qubit
operations on the logical qubits, using suitably timed conjugating operations
derived from analysis of the commutator algebra.  The timing of the operations is
seen to be crucial in allowing simplification of the gate sequences for the
generalized Hamiltonian to forms similar to that derived previously for the
symmetric (XY) anisotropic exchange Hamiltonian.  The total number of operations needed
for a controlled-Z gate up to local transformations is five.  A scalable architecture
is proposed.

\end{abstract} 

\maketitle

\section{Introduction}
\label{sec:intro}

Quantum computation \cite{Gruska:99,Nielsen:00} is known to be universal as long 
as arbitrary single-qubit and non-local (entangling) two-qubit unitary 
operations can be applied in an arbitrarily structured sequence (a quantum 
circuit) \cite{DiVincenzo:95,Barenco:95}.  These operations are generated by 
Hamiltonian control fields, which may be either directly related to interactions 
intrinsic to the physical system of qubits, or may be imposed as additional, 
external, fields.  Universality is thus determined fundamentally by the physical 
structure of the qubit implementation and by control of this.  Early studies of 
universality sought on the one hand to establish specific examples of universal 
gate sets \cite{deutsch:89}, and on the other hand to determine whether there were any 
restrictions on the underlying interactions.  This was nicely summarized by the 
pleasing result that ``almost any'', i.e., a generic two-body interaction could 
provide universal operations \cite{lloyd:95,deutsch:95}.  
Now that experimental efforts to implement 
quantum computation are being undertaken, the question of just which 
interactions to use in a particular physical implementation becomes relevant.  
In particular, there is often a significant distinction between controlling an 
interaction that is intrinsic to the system, and introducing a new interaction 
with an external control field.  In many of the physical implementations that 
have been suggested to date, the inherent physical interactions do not suffice 
to generate the universal set of quantum computing operations over physical 
qubits and must be supplemented by such additional, external Hamiltonian terms.  
This may introduce demanding nanoscale engineering constraints as well as 
additional unwanted sources of decoherence.  Consequently, the question of 
whether and how we can use a particular physical system containing some very 
specific, non-generic interactions, for universal quantum computation has become 
very relevant, now that experimental efforts to implement small scale quantum 
logic are underway.

Recently, Whaley and coworkers \cite{Bacon:01,Kempe:01a,Kempe:ph,Kempe:IP} 
have established a new paradigm of ``encoded universality'', within which the
limitations of non-generic intrinsic physical interactions can be overcome by a 
suitable encoding of the states to be used for quantum logic into a subspace of
the system Hilbert space that is derived from two or more
physical qubits.  This is achieved
with a general algebraic approach that allows the appropriate 
encoding to be determined from analysis of the properties of Lie algebras 
that are generated by the interaction Hamilitonian.   To date, this general algebraic 
encoded universality approach has successfully been applied to two kinds of 
interactions, both of which are variants of the two-particle exchange interaction.  The 
isotropic (Heisenberg) exchange interaction, i.e., $J^x = J^y = J^z \equiv J$
(for meaning of the symbols $J$, see (\ref{Eq:Hamiltonian})), 
was proved universal on encodings of three physical qubits
and higher \cite{Bacon:01,Kempe:01a}. Efficient solutions were found for the smallest 
encoding, when the logical qubit is encoded into 
three physical qubits and the interaction can be implemented between any pair of 
neighboring physical qubits (arranged in a chain or two-dimensional 
lattice) \cite{DiVincenzo:00a}.  
More recently, the symmetric anisotropic exchange, $J^x = J^y \neq J^z = 0$ has 
also been demonstrated to be universal \cite{Kempe:ph,Kempe:IP}.  Like the 
Heisenberg case, the smallest encoding here is also three physical qubits.  In the 
case of the symmetric anisotropic exchange, it was shown that this could be used 
to form either an encoded qubit or an encoded qutrit \cite{Kempe:IP2}.  

These instances of encoded universality derived from a single physical 
interaction are to be distinguished from related results of Levy \cite{Levy:01} and of Wu and Lidar 
\cite{Lidar:01a, Lidar:01b, Lidar:01c} for exchange interactions that are 
necessarily supplemented by an additional single-qubit energy spectrum, 
$\sigma^{z}$.  With this additional interaction, whether imposed statically or 
coupled to an external control field as in the earlier paradigm of controllable 
one- and two-qubit interactions, Wu and Lidar found a two-qubit encoding that 
was universal.  These authors have applied 
this approach incorporating the additional $\sigma^{z}$ interaction
to more general situations described by Hamiltonians containing exchange 
interactions with coupling coefficients $J^{\alpha}_{ij}$ having 
different values \cite{comm:1}).  

The significance of these encoded universality schemes for quantum computation 
lies in the fact that they
require active manipulation only of two-particle exchange interactions, 
and hence can be generically referred to ``exchange-only computation''.  
They are closely related to numerous  proposals for quantum 
computation in solid state systems in which the exchange interaction is a common
feature. These can be summarized as 
follows.
(I) The case of isotropic exchange, $J^{\alpha} = J$ for $\alpha = x,y,z$, is 
represented by spin-coupled quantum dots \cite{Loss:98,Burkard:99,Hu:00} and 
by donor atom nuclear/electron spins \cite{Kane:98}. (II) The symmetric anisotropic 
exchange, $J^{\alpha} = J$ for $\alpha = x,y$, includes quantum dots in cavities 
\cite{Imamoglu:99}, atoms in cavities \cite{Zheng:00}, exciton coupled quantum 
dots \cite{Quiroga:99} and quantum Hall systems \cite{Mozyrsky:01}. 
(III) A more general 
anisotropic exchange, $J^{\alpha} = J \neq J^z \neq 0$ for $\alpha = x,y$, is 
represented by the proposal to use electrons on helium as qubits 
\cite{Platzman:99}.  
While the general two-body 
exchange Hamiltonian $H_{ij} = \sum_{^{\alpha}}J^{\alpha}_{ij} 
\hat{\sigma}^{\alpha}_i {\hat{\sigma}}^{\alpha}_j$, where $\alpha = x, y$ and $z$ 
and ${\sigma}^{\alpha}$ are the Pauli matrices, 
is applicable to these theoretical solid state proposals, we note that this exchange
Hamiltonian does not 
contain any cross product terms ${\hat{\sigma}}^{\alpha}_{i} 
{\hat{\sigma}}^{\beta}_{j}$, ${\alpha} \neq {\beta}$ which may result for 
instance from the Dzyaloshinski-Moriya term of the spin-orbit interaction 
\cite{Dzyaloshinski:58,Moriya:60}.

In this paper we develop the encoded universality representation for the generalized anisotropic 
interaction that results from allowing asymmetry in the
exchange tensor, {\em i.e.}, $J_z = 0$, $J^x \neq J^y$, and that also incorporates additional 
cross-terms, 
${\hat{\sigma}}^{\alpha}_{i} {\hat{\sigma}}^{\beta}_{j}$, ${\alpha} \neq 
{\beta}$, in the Hamiltonian. This interaction becomes relevant to some of the recent 
proposals for solid state implementation of quantum computation when 
additional physical effects such as symmetry breaking perturbations 
\cite{Lidar:01c}, originating {\em e.g.}, from surface and interface effects, 
spin-orbit coupling 
\cite{Dzyaloshinski:58,Moriya:60}, dipole-dipole coupling in the spin-spin 
interaction, and anisotropy in exciton exchange interaction in quantum dots 
\cite{Dzhioev:97, Dzhioev:98,Paillard:01} are taken into account.
The asymmetric anisotropic interaction including these cross-terms is also part of a more 
general model 
considered by Terhal and DiVincenzo \cite{Terhal:01} in the framework of 
fermionic quantum computation \cite{Bravyi:00,Valiant:01}.

The structure of the remainder of this paper is as follows.
In Sections \ref{sec:hamiltonian} and \ref{sec:algebra} we use the algebraic approach developed
by Kempe {\it et al.} \cite{Kempe:01a,Kempe:ph,Kempe:IP}, 
to establish code spaces for universal encoding of the generalized anisotropic exchange
interaction.  These are found to be two 
four-dimensional subspaces of the original Hilbert space for three physical
qubits, characterized by the 
parity of the bit string (even or odd number of logical values 1). A logical qubit
can be formed within any of these four-dimensional code spaces.  In Section \ref{sec:onequbit} 
we consider how to perform single-qubit operations on these encoded qubits. It is shown 
that the asymmetry of the anisotropic interaction is removed via the commutation
relations of two-body interaction Hamiltonians within this subspace.
The $\bar{\sigma}^y$ operation is performed on the logical qubits via the commutation
relations,
while the interaction Hamiltonian 
generates the encoded $\bar{\sigma}^x$, sufficing to generate the entire logical qubit 
su(2) algebra.
In Section \ref{sec:twoqubit} we show how to construct logical two-qubit gates, using
the example of the controlled-Z gate (often referred to simply as C(Z)), 
by making use of a relation between the commutation  relations  and conjugation.
We find that the resulting 
conjugating sequences are - up to the duration of conjugating operations - 
equivalent to that obtained previously for the symmetric anisotropic exchange 
\cite{Kempe:IP2}, and that these results are not affected by the presence of cross-terms
in the generalized exchange interaction. Section \ref{sec:implement} discusses 
implementation issues, including analysis of the efficiency of the resulting
gate sequences, and architectural possibilities. 
We conclude with a brief summary in Section \ref{sec:conclusion}. 

\section{Generalized anisotropic exchange Interaction}
\label{sec:hamiltonian} 
 
\subsection{Asymmetric Anisotropic Exchange Interaction} 
 
The anisotropic exchange interaction between two physical qubits, $i$ and $j$,  
is described by the following Hamiltonian operator, 
 
\be  
{\bf \hat{H}}_{ij} = \frac{1}{2}\sum_{\alpha = x,y} J_{ij}^{\alpha}  
\sigma_{i}^{\alpha} \sigma_{j}^{\alpha} = \frac{1}{2} (  
J_{ij}^{x}\sigma_{i}^{x}\sigma_{j}^{x} + J_{ij}^{y}\sigma_{i}^{y}\sigma_{j}^{y}  
),
\ee 
where $J_{ij}^{\alpha}$ is the coupling strength between the qubits, the upper  
index $\alpha$ corresponds to either the $xx$ or $yy$ term of the exchange  
interaction, and $\sigma^{\alpha}$ are the Pauli matrices. If both coupling  
strengths are identical, the Hamiltonian describes the symmetric anisotropic  
interaction often referred to as the XY model. The encoded universality  
\cite{DiVincenzo:00a,Kempe:01a,Kempe:ph} for this case was studied by Kempe {\it  
et al.} \cite{Kempe:ph,Kempe:IP,Kempe:IP2}. 
The asymmetric anisotropic exchange interaction is defined when $J_{ij}^{x}  
\neq J_{ij}^{y}$. This can be reexpressed as a sum of symmetric $(s)$ and  
antisymmetric $(a)$ terms, 
\be \label{Eq:Ham} 
{\bf \hat{H}}_{ij} = {\bf \hat{H}}_{ij}^{s} + {\bf \hat{H}}_{ij}^{a} =  
\frac{1}{2}[J_{ij}^{s}(\sigma_{i}^{x}\sigma_{j}^{x} +  
\sigma_{i}^{y}\sigma_{j}^{y}) + J_{ij}^{a}(\sigma_{i}^{x}\sigma_{j}^{x} -  
\sigma_{i}^{y}\sigma_{j}^{y})],
\ee 
where $J_{ij}^{s} = \frac{1}{2}(J_{ij}^{x} + J_{ij}^{y})$ and $J_{ij}^{a} =  
\frac{1}{2}(J_{ij}^{x} - J_{ij}^{y})$. This asymmetric anisotropic  
Hamiltonian can be seen to split into two distinct parts that act on orthogonal  
two-dimensional  sectors of the four-dimensional Hilbert space if 
the symmetric term is reexpressed as proportional to
$(\sigma^+_{i}\sigma^-_{j} + \sigma^-_{i}\sigma^+_{j})$, and
the antisymmetric component as proportional to
$(\sigma^+_{i}\sigma^+_{j} + \sigma^-_{i}\sigma^-_{j})$, where 
$\sigma^+ = (\sigma^x + i \sigma^y)/2$ and $\sigma^- = (\sigma^x - i \sigma^y)/2$ are  
raising and lowering operators of the system.
These sectors are characterized by the parity of the bit string which refers to
even or odd occupation number defined as the number of 1's in the bit string 
In particular, the symmetric term ${\bf \hat{H}}_{ij}^{s}$ operates  
in the subspace spanned by $\{|01\ra,|10\ra\}$, and the antisymmetric term ${\bf  
\hat{H}}_{ij}^{a}$ in the subspace spanned by $\{|00\ra,|11\ra\}$. 
We explicitly point out that the symmetric term preserves 
the occupation number, while the antisymmetric changes this occupation number by two.
If these pairs of two-particle states are taken to form logical qubit states, 
then it is easily verified that both ${\bf \hat{H}}_{ij}^{s}$ and ${\bf \hat{H}}_{ij}^{a}$ 
act as ${\sigma}^x$ on these pairs of two-qubit states. 

The origin of the asymmetry in the anisotropic interaction can be  
understood as a consequence of energy non-conserving terms in the system  
Hamiltonian. Since the antisymmetric term is proportional to 
$(\sigma^+_{i}\sigma^+_{j} + \sigma^-_{i}\sigma^-_{j})$, it 
represents an energy non-conserving process similar to the anti-rotating wave terms 
arising in the interaction of a two-level system with semiclasical radiation, but happening  
now in a correlated way on both coupled physical qubits. 
We may assume that asymmetry in the anisotropic exchange interaction between  
physical systems is a consequence of the system complexity when numerous  
mechanisms of mutual coupling take place simultaneously. 
An example of similar symmetry breaking in the  
case of the isotropic (Heisenberg) exchange interaction between quantum dots 
derives from the spin-orbit or a, usually weaker, dipole-dipole coupling.  
 
\subsection{Cross-Product Terms} 
 
In general, it has been recognized recently that anisotropy in the exchange interaction  
may be accompanied by cross-product  
terms in the two-body Hamiltonian \cite{Terhal:01}. These can arise, for instance,  
from spin-orbit coupling \cite{Burkard:01} as noted above. 
The total interaction can then be described as follows, 
 
\be \label{Eq:Hamiltonian}
{\bf \hat{H}}_{ij} = \frac{1}{2} \sum_{\alpha = x,y} J_{ij}^{\alpha}  
\sigma_{i}^{\alpha} \sigma_{j}^{\alpha} + \frac{1}{2} \sum_{\alpha \neq \beta =  
x,y} J_{ij}^{\alpha\beta} \sigma_{i}^{\alpha} \sigma_{j}^{\beta} 
\ee 
 
This Hamiltonian can be reexpressed in a form that emphasizes the effect of its  
various terms on subspaces of different parity (the upper index $s$ for odd and  
$a$ for even): 
 
\ba \label{Eq:TotHam} 
{\bf \hat{H}}_{ij} & = & {\bf \hat{H}}_{ij}^{s} + {\bf \hat{H}}_{ij}^{a} + {\bf  
\hat{h}}_{ij}^{s} + {\bf \hat{h}}_{ij}^{a} \nonumber \\ 
& = & \frac{1}{2}[J_{ij}^{s}(\sigma_{i}^{x}\sigma_{j}^{x} +  
\sigma_{i}^{y}\sigma_{j}^{y}) + J_{ij}^{a}(\sigma_{i}^{x}\sigma_{j}^{x} -  
\sigma_{i}^{y}\sigma_{j}^{y})] +  
\frac{1}{2}[K_{ij}^{s}(\sigma_{i}^{x}\sigma_{j}^{y} -  
\sigma_{i}^{y}\sigma_{j}^{x}) + K_{ij}^{a}(\sigma_{i}^{x}\sigma_{j}^{y} +  
\sigma_{i}^{y}\sigma_{j}^{x})], 
\ea 
where $K_{ij}^{s} = \frac{1}{2} (J_{ij}^{xy} + J_{ij}^{yx})$ and $K_{ij}^{a} =  
\frac{1}{2} (J_{ij}^{xy} - J_{ij}^{yx})$. We note that the cross-product 
terms ${\bf \hat{h}}_{ij}^{s}$  
and ${\bf \hat{h}}_{ij}^{a}$ act on  the  
subspace spanned by the basis states $\{|01\ra,|10\ra\}$ and  
$\{|00\ra,|11\ra\}$, respectively.  Both terms are seen to 
act as a $\sigma^y$ operation on these states.  
 
These subspaces are seen to be spanned  
by basis sets characterized by the bit-string parity  
${\mathcal{B}}^s = \{|01\ra,|10\ra\}$ and ${\mathcal{B}}^a = \{|00\ra,|11\ra\}$.  
The action of the total Hamiltonian is simultaneous in both subspaces.  
In particular, the symmetric  
component of the interaction (indexed $s$) acts only in ${\mathcal{B}}^s$, and  
the antisymmetric part (indexed $a$) only in ${\mathcal{B}}^a$. In each of the two subspaces  
the interaction is characterized by the expression $J_{ij}^{k}  
\sigma^x_{{\mathcal{B}}^{k}} + K_{ij}^{k} \sigma^y_{{\mathcal{B}}^{k}}$, where  
$k$ is either $s$ or $a$. This can be reformulated as ${\tilde{J}}_{ij}^{k}  
\sigma^+_{{\mathcal{B}}^{k}} + {\tilde{J}}_{ij}^{k *} \sigma^-_{{\mathcal{B}}^{k}}$, where the effective coupling now becomes a complex number,  
 
\be \label{Eq:ComplexJ}
{\tilde{J}}_{ij}^{k} = J_{ij}^{k} - i K_{ij}^{k}. 
\ee 
The operators  
${\sigma}^x_{{\mathcal{B}}^{k}},{\sigma}^y_{{\mathcal{B}}^{k}},{\sigma}^+_{{\mathcal{B}}^{k}}$ and $\sigma^-_{{\mathcal{B}}^{k}}$ now apply to the pairs of  
states within any of the two-dimensional subspaces  ${\mathcal{B}}^s$ or ${\mathcal{B}}^a$.  
 
\section{Algebraic Aspects of the Interaction} 
\label{sec:algebra} 
 
The set of asymmetric anisotropic exchange Hamiltonians between neighboring  
physical qubits, given by (\ref{Eq:Ham}), $\{{\bf \hat{H}}_{ij}, 1 \leq i <  
j \leq n \}$, generates the Lie algebra ${\mathcal{L}}$, where $n$ is the total  
number of physical qubits. We consider first the Hamiltonian without cross terms. 
The effect of the cross-product terms will be considered below. We follow here the  
algebraic approach due to Kempe et al. \cite{Kempe:01a,Kempe:ph,Kempe:IP} and first study  
the properties of the algebra commutant ${\mathcal{L}}'$.  Our goal is to  
identify suitable encoding of quantum information and so we do not here exploit the  
potential of the algebraic approach in providing a constructive proof for such  
encodings for general $n$. Since the symmetric anisotropic interaction has been  
proved to be universal over three-qubit and higher encodings 
\cite{Kempe:ph,Kempe:IP,Kempe:IP2}, we shall examine here only the minimal 
case where $n=3$  
to see whether analogous results hold for asymmetric anisotropic exchange. 
 
We identify two operators as the elements of the commutant:

\be 
{\bf \hat{Z}} = \bigotimes_{k=1}^{n}\sigma_{k}^{z}, ~~{\bf \hat{X}} =  
\bigotimes_{k=1}^{n}\sigma_{k}^{x}.
\ee 
Both of these operators commute for even $n$, and anticommute for odd $n$. 
We remark that ${\bf \hat{Z}}$ is a parity operator. This commutes with 
the generalized anisotropic exchange Hamiltonian which preserves the parity.
We point out that in the case of the symmetric anisotropic (XY) interaction, the commutant  
becomes larger, represented now by the operators ${\bf \hat{X}}$ and ${\bf  
\hat{S}}_z = \bigoplus_{k=1}^{n} \sigma^z_k$~\cite{Kempe:ph}.  
In the case of the isotropic exchange  
interaction these are further expanded to ${\bf \hat{S}}_x$ and ${\bf \hat{S}}_z$  
~\cite{Kempe:IP}.  In the present work we focus on the  
universality properties and do not address the decoherence-free aspects of the 
encoding. Also, we note that including the cross-product terms  
into the Hamiltonian operator changes the commutant structure, since ${\bf \hat{X}}$  
is no longer an element of the resulting commutant.  
We show explicitly in Section~\ref{sec:onequbit} that the proposed codes derived from the algebraic  
analysis without 
cross terms are nevertheless also universal for the general case including the  
cross-product terms. The remainder of this Section will therefore continue to 
deal with the algebraic analysis for (\ref{Eq:Ham}) alone. 
 
We assume that the algebra ${\mathcal{M}}$ generated from these operators by  
linear combination and multiplication is identical to the commutant  
${\mathcal{L}}'$. Then, the splitting of ${\mathcal{M}}$ into the irreducible  
representations $J \in \mathcal{J}$, 
 
\be 
{\mathcal{L}'} = {\mathcal{M}} = \bigoplus_{J \in {\mathcal{J}}}  
{\mathcal{I}}_{n_J} \otimes M({\CC}^{d_J}) 
\ee 
translates into the structure of the irreducible representations of the Lie  
algebra generated by the Hamiltonian operators 
\be 
{\mathcal{L}} \cong \bigoplus_{J \in {\mathcal{J}}} {\mathcal{L}}_{J}(n_J)  
\otimes {\mathcal{I}}_{d_J} 
\ee 
over the Hilbert space 
\be 
{\mathcal{H}} \sim \sum_{J \in {\mathcal{J}}} {\CC}^{n_J} \otimes {\CC}^{d_J}, 
\ee 
where $n_J$ and $d_J$ are the dimension and degeneracy of irreducible  
representation $J$, respectively. 
 
For the case of three physical qubits, $n = 3$ (note that $N=2^n$), the operators ${\bf  
\hat{X}}$ and ${\bf \hat{Z}}$ are 8x8 and possess a block diagonal structure of four 2x2  
blocks  
\ba 
{\bf \hat{X}}_{\mathcal{B}} &=& \bigoplus_{k=1}^{N/2} \sigma_k^x \nonumber \\ 
{\bf \hat{Z}}_{\mathcal{B}} &=& \bigoplus_{k=1}^{N/2} \sigma_k^z, 
\ea 
when expressed in the basis set ${\mathcal {B}}$ obtained by a suitable permutation of the  
standard basis: 
 
\be \label{Eq:Permut} 
{\mathcal{B}} =  
\{|000\ra,|111\ra,|110\ra,|001\ra,|101\ra,|010\ra,|011\ra,|100\ra\}. 
\ee 
 
The commutation relation taken over these two operators generates ${\bf  
\hat{Y}}_{\mathcal{B}} = \bigoplus_{k=1}^{N/2} \sigma_k^y$, and hence the  
complete su(2) algebra over the 2x2 blocks.  
 
The algebra ${\mathcal{M}}$ is now expressed as the tensor product 
${\mathcal{I}}_{4} \otimes M({\CC}^{2})$. Consequently, its commutant  
${\mathcal{M}}'$ - associated with the Lie algebra ${\mathcal{L}}$ by our 
assumption that ${\mathcal{M}}$ is identical to the Lie algebra commutant  
${\mathcal{L}}'$ - splits as $M({\CC}^{4}) \otimes {\mathcal{I}}_{2}$. 
The Hilbert space of this system splits accordingly into two four dimensional  
subspaces, ${\mathcal{H}}^8 = {\mathcal{H}}^4 \oplus {\mathcal{H}}^4$, which are  
characterized by different bit-string parities. As expected, these subspaces are  
not mixed by the interaction Hamiltonian (\ref{Eq:Ham}), which preserves the bit-string parity.  
The four-dimensional subspaces thus define two independent codes that are  
spanned by the following two sets of code words: 
 
\ba \label{Eq:Codes} 
(I)&\{|000\ra,|110\ra,|101\ra,|011\ra\}& \nonumber \\ 
(II)&\{|111\ra,|001\ra,|010\ra,|100\ra\}& 
\ea 
These states are used as convenient basis sets for representing (\ref{Eq:Ham}).
In the following they will be referred to the order above, {\it i.e.} states 1, 2, 3, and 4
reading from left to right. 
Before we define the qubit  encoding onto these subspaces, we first examine 
the effect of the asymmetric anisotropic  exchange  
interaction Hamiltonian on the code words.

\section{Single-Qubit Operations} 
\label{sec:onequbit} 
 
\subsection{Asymmetric Anisotropic Exchange} 
 
As shown above, the symmetric and antisymmetric component of the exchange Hamiltonian  
(\ref{Eq:Ham}) act simultaneously on two orthogonal two-dimensional subspaces  spanned 
respectively by ${\mathcal{B}}^s = \{|01\ra,|10\ra\}$ and  ${\mathcal{B}}^a = \{|00\ra,|11\ra\}$.  

We now apply this Hamiltonian to the pairs of physical qubits 1-2,1-3, and 2-3, 
in the three qubit codes given by (\ref{Eq:Codes}). We emphasize that the  
effect of this interaction is the same for both codes, {\em i.e.}, for ($I$) and for ($II$). 
In  fact, the Hamiltonian acts simultaneously and identically on both subspaces ${\mathcal{H}}^4$,  
without mixing them, and it can therefore be expressed as a direct sum of two 4x4 matrices 
in the basis of the codes ($I$) and ($II$).  We now analyze the action of the Hamiltonian on these 
codes.  In the code basis (\ref{Eq:Codes}), the effect of the asymmetric anisotropic 
exchange interaction, schematically summarized in Figure~\ref{fig:H_action},  
applied  to any pair of these qubits possess the following forms:
\ba \label{Eq:Hmat} 
H_{12} =  
\left( 
\begin{array}{cccc} 
0 & J^a & 0 & 0 \\ 
J^a & 0 & 0 & 0 \\ 
0 & 0 & 0 & J^s \\ 
0 & 0 & J^s & 0 
\end{array} \right), \quad 
H_{13} =  
\left( 
\begin{array}{cccc} 
0 & 0 & J^a & 0 \\ 
0 & 0 & 0 & J^s \\ 
J^a & 0 & 0 & 0 \\ 
0 & J^s & 0 & 0 
\end{array} \right), \quad 
H_{23} =  
\left( 
\begin{array}{cccc} 
0 & 0 & 0 & J^a \\ 
0 & 0 & J^s & 0 \\ 
0 & J^s & 0 & 0 \\ 
J^a & 0 & 0 & 0 
\end{array} \right). 
\ea 
Here the lower index indicates between which physical qubits,  $i$ and $j$, 
the interaction is turned on, and
$J^s = J_{ij}^s/2$, $J^a = J_{ij}^a/2$ are the coupling strengths for 
the symmetric and antisymmetric parts respectively of (\ref{Eq:Ham}), for physical qubits $i$  
and $j$. 
For  instance, for a triangular arrangement (shown in Figure~\ref{fig:triangle} bellow) 
the  coupling strengths $J^s$ and $J^a$ are the same in each of  
the Hamiltonian matrices (\ref{Eq:Hmat}), since they are derived from equivalent  
nearest neighbor interactions.  This is necessary for the elimination of the antisymmetric  
component that is accomplished below via use of commutation relations, and it therefore affects 
the architecture of a potential qubit array. This aspect is discussed further in   
Section~\ref{sec:architect} below.  
In order to clarify the notation, we provide an example, noting that, 
{\it e.g.}, the matrix  
$H_{12}$ represents the coupling between the physical qubits 1 and 2  
which simultaneously transforms the logical qubits 1 and 2 (in  
each of the code spaces ($I$) or ($II$)) via the antisymmetric component of the  
Hamiltonian ($J^a$), and the logical qubits 3 and 4 through its symmetric component  
($J^s$). 
 
Let us now consider the action of these three matrices and of their commutators.   
We start with $H_{12} = (J^a \sigma^x) \oplus (J^s \sigma^x)$. 
From  (\ref{Eq:Hmat}) it is evident that the symmetric component of $H_{12}$, 
{\em i.e.}, the lower right 2x2 block, acts as  a ${\sigma}_{34}^x$ operation 
over the code words 3 and 4 from (\ref{Eq:Codes}), {\it i.e.}, the states  
$|101\ra$ and $|011\ra$ from ($I$), with coupling strength $J^s$. It has the same effect  
over the states $|010\ra$ and $|100\ra$ from ($II$), {\it i.e.} it acts as encoded 
${\bar{\sigma}}^x$ on the states in both (I) and (II).
The antisymmetric component of $H_{12}$, which is 
the top left 2x2 block of this matrix, acts on the other  
two orthogonal states from the code, namely on $|000\ra$ and $|110\ra$ from ($I$),  
or on $|111\ra$ and $|001\ra$ from ($II$). 
This also results in an encoded ${\bar{\sigma}}^x$ operation 
but with coupling strength $J^a$. 
This is the first element required for an encoded SU(2) operation. 
The effect of these interactions is summarized schematically in Figure~\ref{fig:H_action}.
 
The second element is an encoded $\bar{\sigma}^y$ operation. These operations 
are generated through the commutator  
of a pair of Hamiltonian operators from (\ref{Eq:Hmat}). For instance,  
taking the commutator of interactions between physical qubits 1-3 and qubits 2-3 
yields $[H_{13},H_{23}] = i [(J^{a})^2-(J^{s})^2] \sigma_{34}^y$, where  
$\sigma_{34}^y$ acts exclusively on the states $|101\ra$ and $|011\ra$. Since all  
other elements of the resulting 4x4 matrix are equal to zero, this commutation  
relation results exclusively in an encoded $\bar{\sigma}^y$ operator between the code  
words 3 and 4, {\em i.e.}, 
\ba \label{Eq:sigmay} 
[H_{13},H_{23}]  = i  
\left( 
\begin{array}{cccc} 
0 & 0 & 0 & 0 \\ 
0 & 0 & 0 & 0 \\ 
0 & 0 & 0 & i[(J^{a})^2-(J^{s})^2] \\ 
0 & 0 & -i[(J^{a})^2-(J^{s})^2] & 0 
\end{array} \right), 
\ea 
 
The third and last element required for an encoded SU(2) operation is 
encoded $\bar{\sigma}^z$. It can be easily verified that these operations are now 
obtainable from a second level commutator, namely  
of the Hamiltonian (\ref{Eq:Hmat}) with the encoded 
$\bar{\sigma}^y$ operations. For example, $[H_{12},\sigma_{34}^y] = i  
2J^s \sigma_{34}^z$.  
 
Together, these three encoded ${\bar{\sigma}}^x$, ${\bar{\sigma}}^y$, and  
${\bar{\sigma}}^z$ operations ensure that any arbitrary SU(2) operation may be 
performed on the encoded qubits. The interactions underlying these operations
and the combinations just described are summarized schematically in Figure \ref{fig:onequbit}. 
We note that the Hamiltonian matrices always act simultaneously on {\it both} sets of  
orthogonal subspaces ($I$) and ($II$).  We can use the encoded operations described 
above to generate additional encoded  
$\bar{\sigma}^x$ operations that do not simultaneously act on the  
orthogonal subspaces from the code, by forming the  
commutator between the $\bar{\sigma}^y$ and  $\bar{\sigma}^z$  
operators.   
 
Analogous sets of operators can be defined starting from the other two  
exchange Hamiltonians, {\em i.e.}, $H_{13}$ and $H_{23}$.  The connections 
resulting from all of Hamiltonian interactions and their commutators are equivalent
in each case to those illustrated in  Figure~\ref{fig:H_action}. 
In total therefore, we have three distinct ways of  
defining the logical qubit from each of the subspaces ($I$) and ($II$), with  
arbitrary SU(2) operations possible on any of these six possible sets of qubits.   
From the 
subspaces ($I$) the possible encodings are $\{|110\ra,|011\ra\}$, or 
$\{|110\ra,|101\ra\}$, or $\{|101\ra,|011\ra\}$. From the subspace  
($II$) the possible qubit  
encodings are $\{|001\ra,|100\ra\}$, and $\{|001\ra,|010\ra\}$, and  
$\{|010\ra,|100\ra\}$.   

Examination of the commutators derived from each of the three different starting exchange 
Hamiltonians shows that the resulting encoded operations, {\it e.g.} (\ref{Eq:sigmay}), 
are in each case 
characterized by zeros in the corresponding locations where 
the antisymmetric terms appear in (\ref{Eq:Hmat}). 
This means that by making use of the commutation relations defined by the interactions of  
two-physical qubits from the three qubit codes, we have completely eliminated the  
effect of that part of the interaction which, as noted before, 
changes the occupation number of the code by two, and 
which corresponds to the antisymmetric component of the Hamiltonian, 
$J^a_{ij}$ (see (\ref{Eq:Ham})).  
Up to a numerical  
factor given by the product of coupling strengths that are accumulated in the course of  
applying the commutation relations, the problem then reduces to that of 
the symmetric anisotropic  
exchange solved previously in \cite{Kempe:IP}. We note that 
in the limit  $J_{ij}^a \to 0$,  the results of Kempe and Whaley 
obtained for encoding into three physical qubits \cite{Kempe:IP,Kempe:IP2} are reconstructed. 
The corresponding three-qubit codes, spanned by  
$\{|110\ra,|101\ra,|011\ra\}$ or $\{|001\ra,|010\ra,|100\ra\}$, define a logical  
qutrit, and the Hamiltonian operators over the code become 3x3 matrices  
generating the complete su(3) Lie algebra \cite{Kempe:IP}. The fact that
SU(2) is a subgroup of SU(3) then further implies existence of 
the truncated qubit representation within 
the three-qubit code that was established in \cite{Kempe:IP2}. In the truncated
qubit representation, one of the physical
qubits is kept constant, and is used merely as an auxiliary element for
generation of the necessary commutation relations. 
 
We also remark that the spin-orbit  
generated anisotropy can be alternatively eliminated to the first order in the spin-orbit
coupling by suitable shaping of the pulsed  interaction between physical 
qubits as recently proposed \cite{Burkard:01,Bonesteel:01}.

\subsection{Cross-Product Terms} 
 
The inclusion of the cross-product terms transforms the Hamiltonian operators  
given by   (\ref{Eq:Hmat})) into hermitian matrices of the same structure whose coupling  
coefficients ${\tilde{J}}^a = {\tilde{J}}_{ij}^a/2$ and ${\tilde{J}}^s = {\tilde{J}}_{ij}^s/2$ 
are now complex (see (\ref{Eq:ComplexJ}). In fact, the situation captured in the Hamiltonian of    
(\ref{Eq:TotHam}) is the most general anisotropic exchange form   
containing asymmetry in all terms including the cross-products. 
It provides a generalization of the usual symmetric anisotropic
exchange referred to as an XY model. 

Under these  
circumstances, application of the commutation relations between the Hamiltonian matrices  
(\ref{Eq:Hmat}) is still capable of generating the su(2) algebra for single qubit operations. 
The result of the commutation relation is again proportional to the $\bar{\sigma^y}$  
operation. For instance, $[\tilde{H}_{13},\tilde{H}_{23}] = i  
(|{\tilde{J}}^a|^2 - |{\tilde{J}}^s|^2) \sigma_{34}^y$.  On the  
other hand, elementary matrix algebra shows that now 
only {\it two} of three possible commutation relations between pairs 
of complex Hamiltonian  matrices (\ref{Eq:Hmat}) of the three-qubit 
code can  eliminate the coupling between states of different occupation number
and thereby generate this encoded $\bar{\sigma^y}$. 
The commutation relation which does not generate this
transformation is $[\tilde{H}_{12},\tilde{H}_{23}]$.
This fact limits which  
two of three possible logical qubit encodings should be considered as universal  
out of the codes listed in (\ref{Eq:Codes}). If only the antisymmetric cross-product term, 
${\bf \hat{h}}_{ij}^{a}$, in (\ref{Eq:TotHam}) is considered ({\it i.e.} $K_{ij}^s = 0$) 
this limitation is removed and all three  
commutation relations can result in cancellation of the antisymmetric component of  
the interaction. 
The other relations generating encoded $\bar{\sigma}^z$ and  $\bar{\sigma}^x$  hold
accordingly.

In conclusion, the commutation relations suffice to completely remove asymmetry  
in the most general anisotropic interaction, including the cross-product terms  
$\sigma^x_i\sigma^y_j$. 
The encodings proposed here provide a direct route to elimination of these terms.  
In Section \ref{sec:implement} we show how to implement the commutation relations 
efficiently.
 
\section{Two-Qubit Operations} 
\label{sec:twoqubit} 

An entangling two-qubit gate - namely the controlled-Z (C(Z)) 
operation - is obtained via the following sequence of encoded $\bar{\sigma}^z$ operations:

\be \label{Eq:cz}
{\bf \hat{U}}_{C(Z)} = e^{i {\bar{\sigma}}_{1}^z \pi / 4} e^{i {\bar{\sigma}}_{2}^z \pi / 4} e^{-i ({\bar{\sigma}}_{1}^z \otimes {\bar{\sigma}}_{2}^z) \pi / 4}.
\ee
The crucial element of this sequence is the last term on the right hand side. 
This is enacted
by applying the encoded  $\bar{\sigma}^z$ operation onto the triplet of 
physical qubits 2-3-4 that connects two logical qubits within the triangular 
architecture (see Figure~\ref{fig:triangle}).  
To illustrate this C(Z) sequence, we focus on 
an  example with the following encoding of logical qubit: $|0_L\rangle = |110\rangle$,
$|1_L\rangle = |011\rangle$. The logical two-qubit configurations 
are then given as

\ba
|0_L0_L\rangle & = & |1\framebox[1.1\width]{{\bf101}}10\rangle \nonumber \\
|0_L1_L\rangle & = & |1\framebox[1.1\width]{{\bf100}}11\rangle \nonumber \\
|1_L0_L\rangle & = & |0\framebox[1.1\width]{{\bf111}}10\rangle \nonumber \\
|1_L1_L\rangle & = & |0\framebox[1.1\width]{{\bf110}}11\rangle,
\ea
where the boxes indicate those physical qubits which are 'bridging' two logical qubits. 
Via commutation relations
of the exchange Hamiltonians between the physical qubits 2-4 and 3-4 within the triangular
architecture (Figure ~\ref{fig:triangle}) we generate the ${\sigma}^y_{2,3}$
operation which, when commuted further with the exchange interaction between 
the qubits 2 and 3, results in the corresponding ${\sigma}^z$ operation.
Turning this ${\sigma}^z$ operation on for the duration $t = \pi /2$  
results in a phase transformation of the states, such that
$|0_L0_L\rangle = |1\framebox[1.1\width]{{\bf101}}10\rangle \to e^{-i\pi/2} |0_L0_L\rangle$ and $|0_L1_L\rangle = |1\framebox[1.1\width]{{\bf100}}11\rangle \to e^{i\pi/2} |0_L1_L\rangle$.  The other two states are not addressed by 
the encoded operation and remain intact. The resulting diagonal transformation
over the logical two-qubit states, characterized by diagonal elements  $\{-i,i,1,1\}$,
has provided the desired entanglement between the logical qubits. 
We emphasize that we needed one double commutator to obtain this transformation.
In order to illustrate that this suffices to generate the controlled-Z operation, we
first apply an encoded $\bar{\sigma}^z$  onto the second logical qubit for 
duration $t = \pi /4$. This further transforms the relative phase relations between
the states of two logical qubits to $\{-i,1,1,-i\}$ (up to an overall phase $e^{i\pi/4}$). 
This result is equivalent to the unitary
transformation  $e^{-i ({\bar{\sigma}}_{1}^z \otimes {\bar{\sigma}}_{2}^z) \pi / 4}$ 
in (\ref{Eq:cz}). This transformation, when supplemented
by the encoded single qubit $\bar{\sigma}^z$ rotations on both logical qubits, results in
the desired controlled-Z operation, C(Z) \cite{Nielsen:00}.  
 
\section{Implementation Issues} 
\label{sec:implement} 

We now turn our attention to practical aspects of implementation of universal quantum 
computation with generalized anisotropic exchange interactions. Our goal is now to translate
the theoretical development of encoded universality with this class of Hamiltonians into 
an appropriate quantum circuit. So far, we have employed the commutation relations between the  
interaction Hamiltonians (\ref{Eq:Hmat} ) to generate an su(2) algebra over a suitably 
selected qubit from one of the code subspaces (\ref{Eq:Codes}). A practical question 
is how to implement the commutation relations. In principle, this
can always be carried out via the Baker-Hausdorff-Campbell operator  
expansion \cite{Kempe:IP}. However this does not necessarily provide the efficiency  
required in practical implementation. A useful approach in the present context is 
based on conjugation by unitary operations considered previously by Kempe {\it et al.}  
\cite{Kempe:ph,Kempe:01a} and by Lidar and Wu \cite{Lidar:01c}. 
Conjugation was recently applied to the case of  symmetric  
anisotropic exchange interactions by Kempe and Whaley \cite{Kempe:IP2}. 
The key observation here was that in the three-qubit encoding of a logical qutrit, 
the complete SU(3) Lie group can  
be obtained through conjugating the evolution operators generated by the  
symmetric anisotropic Hamiltonians over the physical qubits. 

A general feature of conjugation operations that 
we would like to stress in the present context, is
that they can provide the same effect over the encoded qubit as exponentiated commutation 
relations.  The  goal is therefore 
to find a conjugating condition under which this equivalence holds. In  
the present case, the antisymmetric term in the Hamiltonian (\ref{Eq:TotHam}) 
complicates the situation, 
since the conjugating sequence of the unitary evolutions in general  
mixes different states of the code space, and may also result in  leakage
of the encoded qubit population into the orthogonal part of the code subspace. 
We note however that the mixing effect  
of the symmetric and antisymmetric term of the Hamiltonian (\ref{Eq:Ham}) can be  
eliminated by choosing a suitable duration of the exchange interaction. Since  
$[{\bf \hat{H}}_{ij}^{s},{\bf \hat{H}}_{ij}^{a}] = 0$ for any $i$ and $j$, the  
unitary evolution operator generated by the Hamiltonian (\ref{Eq:Ham}) splits  
into a product $U(\tau) = exp(-i {\bf \hat{H}}_{ij}^{s} \tau)exp(-i {\bf  
\hat{H}}_{ij}^{a} \tau)$. For a suitably chosen time duration, one of the terms  
can always be made to
generate the identity from the interaction Lie group, if $J_{ij}^{s} \neq  
J_{ij}^{a}$. At the same time, the effect of the other term can be tuned to  
provide desired transformation. 
 
\subsection{Single qubit operations}

We now illustrate this possibility of turning off the mixing effect of 
the  antisymmetric terms in the evolution operator by judicious choice of conjugation  
operations, with a specific example. 
For instance, the unitary evolution generated by the $\sigma_{34}^y$  
operator, resulting from the commutation relation  
$[H_{13},H_{23}]$, can be obtained from  
the following conjugation:
 
\be \label{Eq:Seq}
{\bf \hat{U}}(\sigma_{34}^y,\phi) = e^{-i \sigma_{34}^y \phi} = e^{i  
H_{13} \theta}  e^{i H_{23} \phi'}  e^{-i H_{13}  
\theta},
\ee 
where $\phi'=\phi/J^s$, and $\theta$ is the time duration satisfying  
simultaneously the two conjugation 
conditions 
\be \label{Eq:Theta}
\theta = 0(mod ~\pi)/J^a = \frac{\pi}{2}(mod ~\pi)/J^s.
\ee 
Due to  the asymmetry of the exchange coupling terms ($J^s \neq J^a$) and to
the unitarity of  the quantum evolution, this condition can easily be fulfilled,
as long as the ratio of $J^s$ and $J^a$ is not a rational number. We note
that rational numbers create a subset of measure zero within the set of real numbers,
and hence it is very unlike that we would meet such a situation in experimental
implementations.

In order to further elucidate the effect of conjugation operations, we focus on analysis
of the conjugating sequence expressed by (\ref{Eq:Seq}). 
The timing condition (\ref{Eq:Theta}) sets the unitary conjugation operator 
into the following matrix form in the code basis (\ref{Eq:Codes}): 

\ba \label{Eq:Decomp}
{\bf \hat{U}}(H_{13},\theta) = e^{i H_{13} \theta} =  
\left( 
\begin{array}{cccc} 
1 & 0 & 0 & 0 \\ 
0 & 0 & 0 & i \\ 
0 & 0 & 1 & 0 \\ 
0 & i & 0 & 0 
\end{array} \right) 
=
\left[
\left( 
\begin{array}{cc} 
1 & 0 \\ 
0 & i 
\end{array} \right)
\oplus
\left( 
\begin{array}{cc} 
1 & 0 \\ 
0 & i 
\end{array} \right) \right]
~{\bf \hat{P}}_{24}
=
({\bf \hat{S}} \oplus {\bf \hat{S}}) ~{\bf \hat{P}}_{24}
.
\ea 
Here ${\bf \hat{P}}_{24}$ is the permutation matrix exchanging the basis states 
$|110\rangle$ and  $|011\rangle$ of the code $(I)$, or
$|001\rangle$ and  $|100\rangle$ of the code $(II)$. The operator ${\bf \hat{S}}$
is an operator inducing the shift of the relative phase by $i$ \cite{Nielsen:00}.
We emphasize that the antisymmetric term of the Hamiltonian $H_{13}$
results in identity, while the symmetric term results in exchange between two code words
that are phase-shifted by $i$.

The identity
${\bf \hat{U}} e^{-i {\bf \hat{H}} \tau} {\bf \hat{U}}^\dagger = e^{-i {\bf \hat{U}} {\bf \hat{H}} {\bf \hat{U}}^\dagger \tau}$ allows us 
to reduce the act of conjugation of the unitary operation to
the conjugation of its generator. Since the Hamiltonian matrices, ({\ref{Eq:Hmat}),
express interactions between different pairs of physical qubits 
within three qubit codewords, they
are related to each other by permutation operations. For instance, the matrix
$H_{12}=(J^a \sigma^x) \oplus (J^s \sigma^x)$ 
can be expressed as ${\bf \hat{P}}_{24} H_{23} {\bf \hat{P}}_{24}$. 
We can now express the desired conjugation in the following form,

\ba \label{Eq:Block}
e^{i H_{13} \theta}  H_{23} e^{-i H_{13} \theta} & = &
({\bf \hat{S}} \oplus {\bf \hat{S}})
{\bf \hat{P}}_{24} ~H_{23}~ {\bf \hat{P}}_{24}
({\bf \hat{S}^\dagger} \oplus {\bf \hat{S}^\dagger}) \nonumber \\
& = & (J^a {\bf \hat{S}} \sigma^x {\bf \hat{S}^\dagger}) \oplus (J^s {\bf \hat{S}} \sigma^x {\bf \hat{S}^\dagger}) \nonumber \\
& = & (J^a \sigma^y) \oplus (J^s \sigma^y),
\ea

Via conjugation, we have now obtained the 2x2 block diagonal matrix
whose blocks are now proportional to the Pauli matrix $\sigma^y$. 
This procedure provides a generalization of the well-known conjugation of the Pauli matrices:
$e^{i \sigma_z \pi/2} e^{i \sigma_x \phi} e^{-i \sigma_z \pi/2} = e^{i \sigma_y \phi}$.
It allows us to generate the full su(2) algebra in each block,
via conjugation with the Hamiltonian $H_{12}$. The block-diagonal structure 
ensures that the antisymmetric component of the interaction does not mixes
with the symmetric one. In contrast to the effect of commutation relations between
the Hamiltonian matrices (\ref{Eq:Hmat}), which eliminate the coupling between
states of different occupation number within the codes (\ref{Eq:Codes}), 
each block in (\ref{Eq:Block}) can now be used for single qubit operations over
the corresponding code states. For the sake of simplicity, we choose the same
qubit encodings as emerged from the commutation relations in Section \ref{sec:onequbit}.
The conjugation can alternatively be formulated to generate the $\sigma^y$ transformations
corresponding to the other two Hamiltonian operators in (\ref{Eq:Hmat}).
 
Development of a conjugating procedure for the case of the general Hamiltonian  
(\ref{Eq:TotHam}), containing the cross-product terms, 
is possible within the same framework. However the relevant timing conditions
have to reflect that the coupling coefficients $\tilde{J}^a$ and $\tilde{J}^s$ may 
now be complex numbers. Just as in the previous case, the goal is to generate the  
desired conjugating unitary transformation by exponentiating the appropriate
general Hamiltonian operator, where the symmetric part
of the interaction leads to exchange between the two coupled code words
phase-shifted by $i$, and the antisymmetric term results in identity 
(see (\ref{Eq:Decomp})).
We first illustrate new timing conditions derived from focusing only on
the antisymmetric term in the generalized anisotropic exchange.

The antisymmetric coupling acts on the state with even bit-string  
parity, ${\mathcal{B}}^a = \{|00\ra,|11\ra\}$. It can be reformulated as the sum  
${J}^a\sigma^x_{{\mathcal{B}}^a} + {K}^a\sigma^y_{{\mathcal{B}}^a}$ where  
$\sigma^x_{{\mathcal{B}}^a}$ and $\sigma^y_{{\mathcal{B}}^a}$ refer only to the  
even parity states. In order to establish the conjugating condition, this  
operator is exponentiated and factorized into the product of three unitary  
operators $e^{-i J^a \sigma^x \Theta} e^{-i K^a \sigma^y \Theta} e^{i J^a K^a  
\sigma^z \Theta /2}$. The condition for attaining the identity is then:
\be \label{Eq:Theta2}
\Theta =  
0(mod ~\pi)/J^a = 0(mod ~\pi)/K^a = 0(mod ~\pi)/(J^a K^a/2). 
\ee
Considering now in addition that the coupling coefficient $J^s$ is complex 
and its imaginary part is also to be  
eliminated, an analogous timing condition can easily be formulated.  
 
The second conjugation needed for $\bar{\sigma}^z$ operations, 
implementing the double commutator (Section \ref{sec:onequbit}), is carried  
out in a similar fashion:
 
\be \label{Eq:SeqZ}
{\bf \hat{U}}(\sigma_{34}^z,\phi) = e^{-i \sigma_{34}^z \phi} = 
e^{i H_{12} \theta} {\bf \hat{U}}(\sigma_{34}^y,\phi)  e^{-i H_{12}  
\theta},
\ee 
Here ${\bf \hat{U}}(\sigma_{34}^y,\phi)$ is the result of 
the first conjugation given by (\ref{Eq:Seq}).
In our example, the condition for the time duration of the second  
conjugating operation generated by $H_{12}$ reads as 
$\theta = 0(mod ~\pi)/J^a = \frac{\pi}{4}(mod ~\pi)/J^s$. 
 
\subsection{Two qubit operations}

We now focus on specific aspects of implementation of the two-qubit gates via
conjugation. The entangling part of the controlled-Z gate,
described in Section \ref{sec:twoqubit} above, is obtained as a conditional effect of 
the ${\sigma}^z$ operation on the physical qubits of both logical qubits 
(on the 'bridging' qubits). 
The conjugation however complicates the situation, due to its antisymmetric component
which affects also the states 
$|1_L0_L\rangle = |0\framebox[1.1\width]{{\bf111}}10\rangle$ and 
$|1_L1_L\rangle = |0\framebox[1.1\width]{{\bf110}}11\rangle$.
However, the effect of the antisymmetric term in the interaction can 
be completely eliminated by imposing an additional timing condition for the conjugated operation.
In our specific example, this operation was generated by  $H_{23}$, and the timing condition
is then given as follows:
\be \label{Eq:Phi}
\phi' = 0(mod ~2 \pi)/J^a = {\phi}(mod ~2 \pi)/J^s.
\ee
We emphasize that this condition has to be satisfied only up to an arbitrary global phase.

\subsection{Efficiency}

The present approach based on conjugation is much more effective than  
application of the Baker-Hausdorff-Campbell formula 
\be
e^{i[{\bf \hat{A}},{\bf \hat{B}}]} = \lim_{n \to \infty} e^{-i {\bf \hat{A}}/\sqrt{n}} e^{i {\bf \hat{B}}/\sqrt{n}} e^{i {\bf \hat{A}}/\sqrt{n}} e^{-i {\bf \hat{B}}/\sqrt{n}}
\ee
whose asymptotic character translates into a sequence of a large number of 
elemetary operations. In contrast, the conjugation requires only
three gates for implementation of the encoded $\bar{\sigma}^y$ operation, emulating
a single commutation relation, and five gates for encoded $\bar{\sigma}^z$, corresponding
to a double commutator.
The entangling two-qubit operation, {\it i.e.} the controlled-Z up to the local
transformations, is based on generating $\bar{\sigma}^z$, and hence requires also
just five discrete gates. This is the same as in the case of the symmetric anisotropic 
interaction studied previously \cite{Kempe:IP2}. The timing conditions, 
expressed in number of gates (\ref{Eq:Theta}) and (\ref{Eq:Theta2}), translate into
a prolonged transformation of the conjugating unitaries. 
It should be pointed out that the duration of the conjugating operation,
given by the ratio of the coupling coefficients $J^a$ and $J^s$ in (\ref{Eq:Theta})
for instance, does  not change if 
larger number of logical  qubits defined with this three-qubit encoding are addressed with
these gate sequences. 
Therefore this approach scales well with size, having only a linear cost in terms of 
computational complexity as the number of encoded qubits increases.  

An alternative to the  
present analytical approach based on conjugation is a numerical optimization of  
gate sequences in order to generate the desired quantum computing operations  
\cite{DiVincenzo:00a}. 
 
\section{Architecture} 
\label{sec:architect} 
 
Since the $\bar{\sigma}^y$ interactions for a given Hamiltonian are defined  
through the commutation relations with the other two available couplings among  
three physical qubits, the most suitable architecture is triangular.
This is summarized in Figure~\ref{fig:onequbit}. An equilateral  
triangular architecture then ensures that $J^s$ and $J^a$ are the same  
within any pair of physical qubits taken from a three qubit code. 
To additionally accomodate also  
two-qubit logical operations, it is convenient to arrange  triangles of physical  
qubits into a linear chain with alternating triangle orientations. 
This layout is shown in Figure~\ref{fig:triangle}. Other layouts, such as
a hexagonal lattice, may also be employed. 

The implementation of commutation relations between exchange Hamiltonains via 
unitary conjugation allows for a number of other architecture structures than equilateral
triangle. The change in the coupling strengths
between physical qubits, which may result from other architectures, 
would be reflected in the timing conditions for conjugating operations discussed above.
In fact, this flexibility is an important aspect of the implementation of 
the Lie algebra of the generalized anisotropic exchange 
via unitary conjugation, because it allows one to relax
the requirement of an equilateral triangular architecture, 
to a lattice of a rectangular or any other structure in order to accomodate 
physical and experimental requirements.

\section{Conclusion} 
\label{sec:conclusion} 
 
In the present work we have demonstrated that encoded universality may be achieved for 
generalizations of the anisotropic exchange interaction that remove the symmetry 
between exchange components acting in the $x$ and $y$ directions, and that also incorporate 
cross-product terms in the Hamiltonian. Using the algebraic approach due to Kempe 
{\it et al.} \cite{Bacon:01,Kempe:01a,Kempe:ph,Kempe:IP}, we find that the Lie algebra generated
by asymmetric anisotropic exchange interaction within encoding into three physical qubits
splits into two irreducible representations that act correspondingly on two
invariant four-dimensional subspaces of the Hilbert space. Their basis sets are used to
define two sets of four code words each. Analysis of  actions of generalized exchange
interactions and their commutation relations within three physical qubits
results in generation of the full su(2) algebra over a single logical qubit.
The most suitable architecture, capturing both the physical properties of the code and
the interactions among its elements, is a chain of equilateral triangles of alternating
orientations. Application of encoded operations within physical qubits
connecting two logical qubits is shown to result in an entangling two-qubit 
operation, namely the controlled-Z.

Implementation issues, related to the efficient implementation of
the commutation relations among exchange interactions, were studied in connection
with the properties of unitary conjugation. 
It was shown that the effect of the commutation relation between
a pair of physical interactions is perfectly mimicked by suitably timed conjugation
of the unitary operations that are generated by these Hamiltonians. 
The timing conditions, explicitly formulated here, result in
significant improvement of implementation efficiency, compared to both 
the asymptotic approach based on the Baker-Hausdorff-Campbell formula
and to recent numerically optimized gate sequences for exchange Hamiltonian \cite{DiVincenzo:00a}. 
The results were found to be valid also in the presence of cross-product terms 
$\sigma^x_i\sigma^y_j$ in the generalized Hamiltonian.
Within the implementation of unitary operations via unitary conjugation, proposed here,
the proposed equilateral triangular architecture may be relaxed according to
the experimental situation. 
 
\begin{acknowledgments}

We thank Julia Kempe for many fruitful discussions and Kenneth Brown for comments on the manuscript.
The effort of the authors is sponsored by the Defense Advanced Research Projects Agency (DARPA)
and the Air Force Laboratory, Air Force Material Command, USAF, under agreement numbers
F30602-01-2-0524 and FDN00014-01-1-0826.
The U.S. Government is authorized to reproduce and distribute reprints for Governmental
purposes notwithstanding any copyright annotation thereon. The views and conclusions
contained herein are those of the authors and should not be interpreted as necessarily 
representing the official policies and endorsements, either expressed or implied, of 
the Defense Advanced Research Projects Agency (DARPA), the Air Force Laboratory,
or the US Government.
 
\end{acknowledgments}

\newpage

\begin{figure}[!h]
\psfig{figure=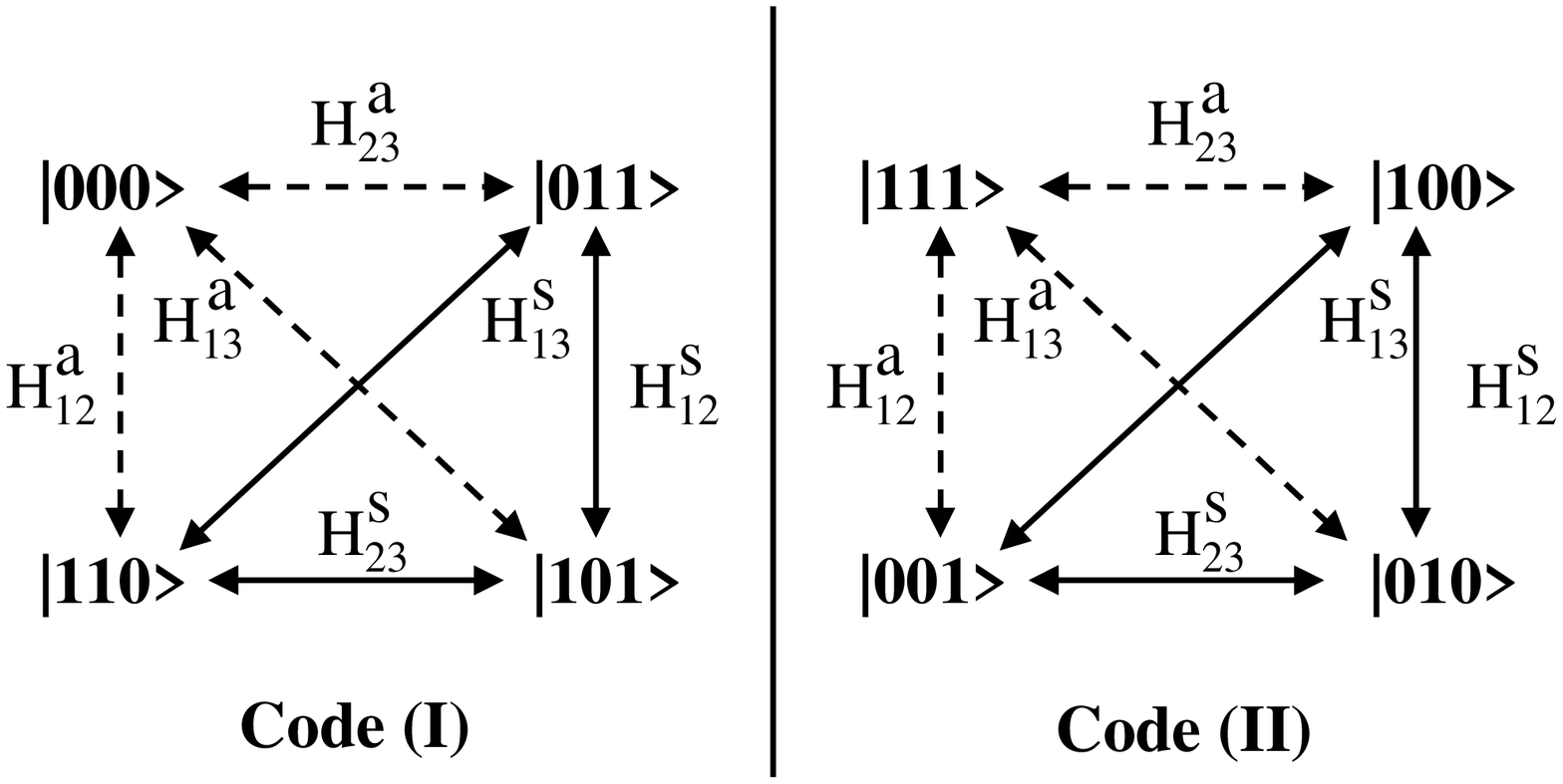,width=1.\textwidth}
\caption{Actions of the asymmetric anisotropic exchange interactions over the three qubit
code spaces. The solid line transforms code words via the symmetric component of 
the Hamiltonian while the dashed line through its antisymmetric part. The earlier
changes the bit-string parity and preserves the occupation number; the latter 
changes the occupation number by two while conserving the parity. Indexes indicate
which physical qubits are coupled.}
\label{fig:H_action}
\end{figure}

\newpage

\begin{figure}[!h]
\psfig{figure=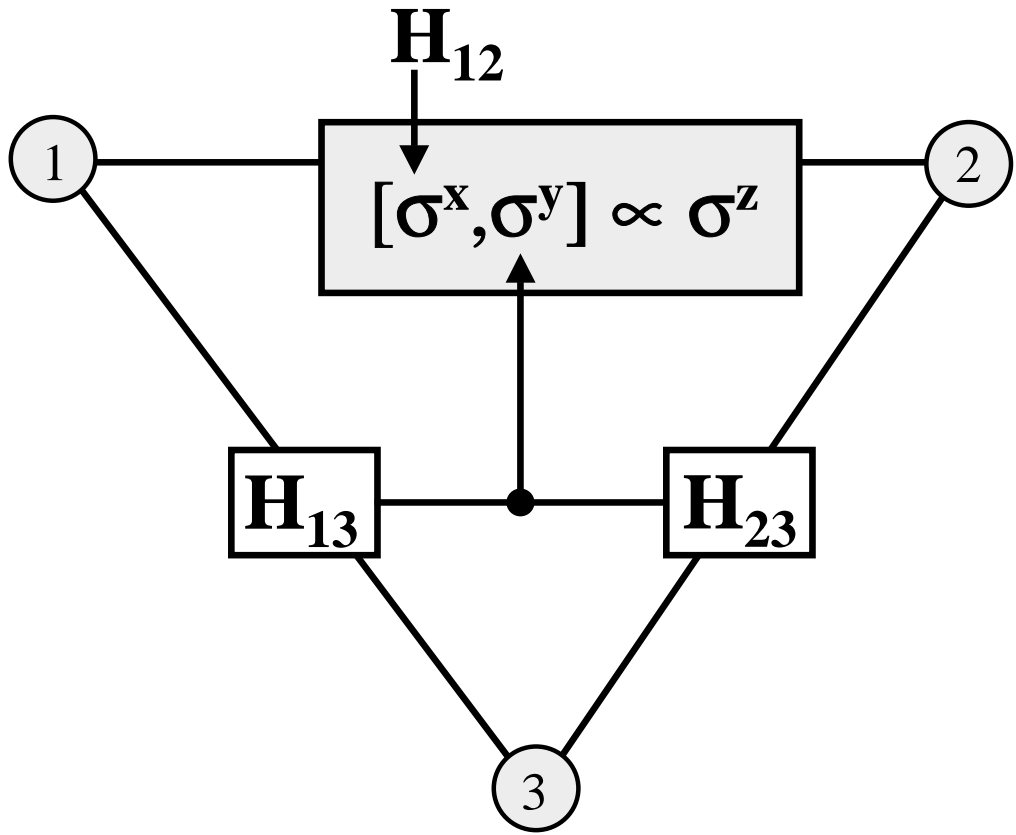,width=1.\textwidth}
\caption{The commutator algebra between the exchange interaction within the three qubit
encoding generates the full su(2) over the encoded logical qubit. The commutation relations
generating this single qubit operations dictate that the appropriate architecture be 
an equilateral triangle.}
\label{fig:onequbit}
\end{figure}

\newpage

\begin{figure}[!h]
\psfig{figure=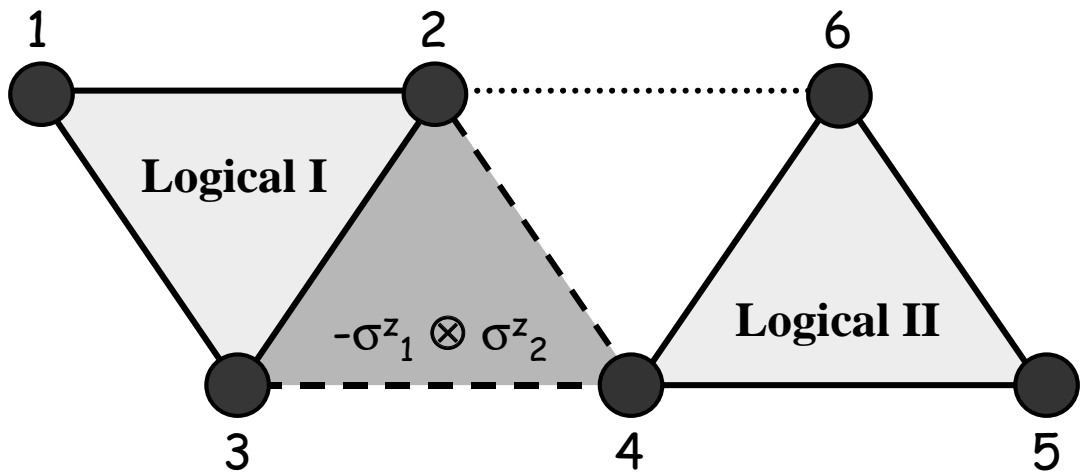,width=1.\textwidth}
\caption{The layout of the scalable architecture. The two qubit entangling operation,
a controlled-Z gate, is implemented using the physical qubits connecting two logical
qubits within the layout, {\it e.g.} physical qubits 2-3-4.}
\label{fig:triangle}
\end{figure}

\newpage

\begin{figure}[!h]
\psfig{figure=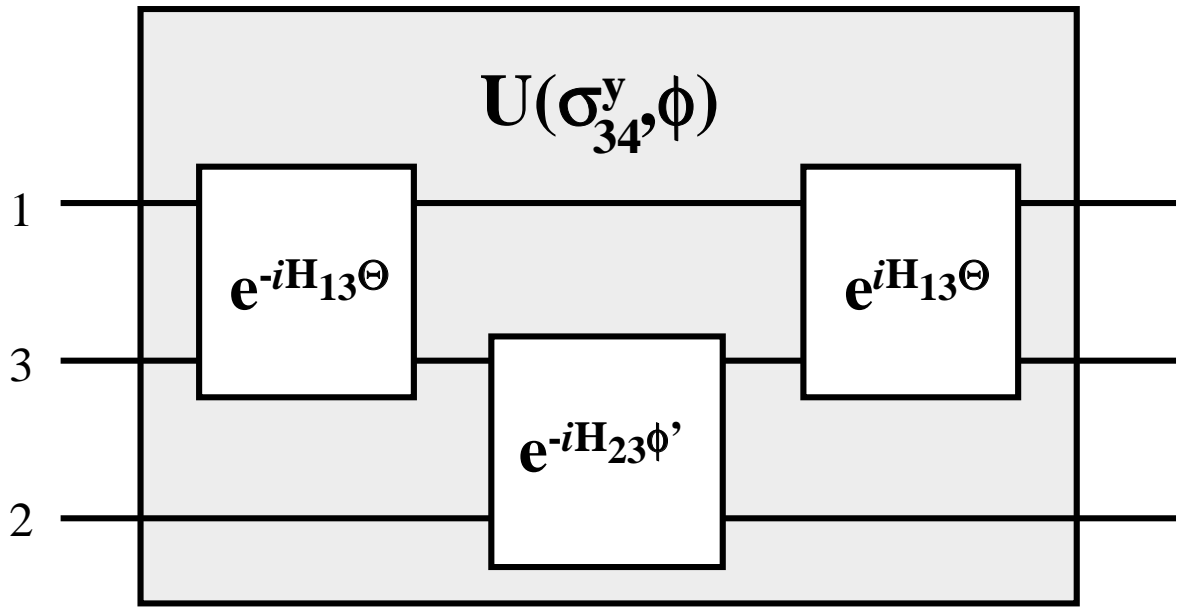,width=1.\textwidth}
\caption{A quantum circuit for generation of the encoded ${\bar \sigma}^y$ operation via
unitary conjugation.}
\label{fig:circuit}
\end{figure}

\end{document}